\documentstyle[pra,aps,psfig,floats,epsf]{revtex}

\newcommand{\be}{\begin{equation}}
\newcommand{\ee}{\end{equation}}
\newcommand{\bea}{\begin{eqnarray}}
\newcommand{\eea}{\end{eqnarray}}
\newcommand{\half}{\mbox{$\textstyle \frac{1}{2}$}}

\newcommand{\shalf}{\mbox{$\textstyle \frac{1}{\sqrt{2}}$}}
\newcommand{\ket}[1]{ | \, #1  \rangle}

\newcommand{\abs}[1]{ | \, #1 \,  |}

\begin{document}
\title{Quantum Key Distribution: from Principles to Practicalities}
\author{Dagmar Bru\ss $^1$\footnote{Present affiliation:
Inst. f\"{u}r Theoret. Physik, Universit\"{a}t Hannover,
Appelstr. 2, D-30167 Hannover, Germany} and Norbert L\"utkenhaus$^2$}
\address{$^1$ISI, Villa Gualino, Viale Settimio Severo 65, I-10133
Torino, Italy}
\address{$^2$Helsinki Institute of Physics, PL 9, FIN-00014
Helsingin yliopisto, Finland}
\date{\today}
\maketitle
\begin{abstract}
We review the main protocols for key distribution based on principles of
quantum mechanics, describing  the general  underlying ideas,
 discussing implementation requirements and 
  pointing out directions of current experiments. 
The issue of security is addressed both from a principal and real-life point of
view. 
\end{abstract}
\pacs{03.67.Dd, 03.67.-a, 03.65.-w, 03.65.-Bz, 42.79.Sz}
\widetext

\section{Principles}
The desire and necessity to transmit secret messages from one person
 to another is probably as old as the capability of human beings to
 communicate.  Cryptography is the art to encode a text in such a way
 that a spy (or eavesdropper) can get as little information as
 possible about it, and only the authorized receiver can decode it
 perfectly. The methods to perform this task have been improved over
 thousands of years.  An important class of  today's schemes  are public-key
 crypto-systems \cite{public}, in which mutually inverse
 transformations are used for encoding and decoding. The instruction
 for encoding is made public, and safety relies on the high complexity
 of the inverse transformation (factorization of large prime
 numbers). In principle this system could be broken, though, by faster
 algorithms (see Shor's algorithm in quantum computation).  \par The
 only crypto-system that has been proven to be safe is  using a random
 key which is only known to the sender and the receiver.  The recipe
 for the sender is to translate the text with a look-up table into a
 sequence of 0's and 1's, e.g., A$\rightarrow 00001$, B $\rightarrow
 00011$, etc., (this translation alone is fairly easy to decipher by
 an enemy) and then to add modulo 2 the random key (a random sequence
 of 0's and 1's), which needs to be of the same length as the message.
 \par The result is that letters which were the same in the original
 message are encoded into completely uncorrelated strings. Only the
 receiver can decode the message by adding again the secret key. This
 method is only safe, though, if the key is used just once, otherwise
 consecutive messages reveal information about the
 messages.\footnote{By adding two messages encoded with the same key
 one obtains the sum of the two original messages. This narrows down
 the possible combinations and reveals a considerable amount of
 information to an eavesdropper.} 
 Therefore this type of protocol is also labeled
 with the key word ``one-time pad'', because in the second 
 world war the key would be written on a
 sheet torn from a pad.  \par Unfortunately, the problem of secrecy is
 hereby only shifted to the problem of distributing the key in a safe
 way to the receiver. In principle, a spy can get hold of the key,
 copy it and send it on to the receiver. This is the point where
 quantum physics enters the stage: if the key distribution makes use
 of quantum states (this is possible in different ways which will be
 explained in detail in the following) the spy cannot measure them
 without disturbing them.  Thus principles of quantum mechanics can
 help to make the key distribution safe. Often this young research
 area is, slightly misleading, also referred to as quantum
 cryptography (for an introduction see \cite{bbe}).  \par In the
 modern communication society there is widespread need of secure
 transmission of secret information (e.g. credit card numbers,
 passwords).  Therefore, a practical realization of these ideas is
 certainly very desirable, and some experimental results have indeed
 already been achieved. We will summarize the occurring problems and
 solutions for some of them and point out the open questions.  \par
 Let us list the main ingredients of quantum mechanics which allow for
 different protocols of secure key distribution - these will be
 explained in more detail in the following chapter:
\begin{itemize}
\item
{{\em non-orthogonal states cannot be distinguished perfectly} \newline
A quantum mechanical 
two-state system  cannot only be in
the state $\ket{0}$ or $\ket{1}$, 
but more  generally in a linear superposition 
$\ket{\psi}=\alpha\ket{0}+\beta\ket{1}$ with 
complex coefficients $\alpha$ and $\beta$ satisfying 
$\abs{\alpha}^2+\abs{\beta}^2=1$.
Due to the laws  of quantum mechanics,
it is impossible to distinguish reliably between 
\bea
\ket{\psi_1}&=&\alpha_1\ket{0}+\beta_1\ket{1}  \ \ \text{and} \nonumber \\
\ket{\psi_2}&=&\alpha_2\ket{0}+\beta_2\ket{1}
\eea
unless  the state overlap is $\langle\psi_1\ket{\psi_2}=0$, 
i.e. the states are orthogonal.}
\item
{{\em no-cloning theorem} \newline
It is impossible, due to linearity and unitarity of quantum mechanics,
 to create perfect copies of an unknown quantum state
 \cite{wootters}. 
 Thus a spy is not able to produce perfect copies of
  a quantum state in transit in order to measure  it, while 
   sending on
  the original.
}
\item
{{\em entanglement (quantum correlation)} \newline
Two or more 
quantum systems can be correlated  or entangled. 
An entangled state cannot be written as a direct product of the subsystems.
The
singlet of two spin-\half-systems is an example of a maximally entangled state:
\be
\ket{\psi^-}=\shalf(\ket{01}-\ket{10}) \ \ \ .
\ee
(The four maximally entangled states of two spin-\half-systems are called
Bell states.)
If a measurement is done on one of these two quantum systems (in any basis),
the result will be 0 or 1 with equal probability. The state of the other system
is anti-correlated, i.e. if the first system collapsed into 0, the second
collapses into 1 and vice versa. Without any measurement, though, 
none of the two
systems 
{\em is} in a
fixed state.
 }
\item
{{\em causality and superposition } \newline
 Causality is {\em not} an ingredient of non-relativistic quantum mechanics. 
 Nevertheless it is mentioned in this list of principles because together with 
 the superposition principle it can be used for secure key distribution: 
 if the two terms of which a superposition consists are sent with a time delay
 relative to each other, such that they are  not causally connected, the
 eavesdropper cannot spy on them.} 
\end{itemize}

\section{Concrete Protocols}
In this chapter we explain different approaches to the task of
establishing a common secret key between two parties. The sender of
the key is usually called Alice and the receiver Bob. Here we will
assume that no enemy (usually called Eve) is present. In chapter
\ref{secure} we will then discuss how a spy can gain some information
on the key.
\par 
We  can distinguish the following main three classes of protocols.
\begin{itemize}
\item [1)]{\em BB84 class:} \newline In 1984 Bennett and Brassard
suggested a quantum cryptographic protocol that relies on the use of
non-orthogonal states \cite{bb84}. It is often referred to as
BB84. There have been several ideas for variations of this protocol
which will for this review be included in the BB84-class.
\begin{itemize}
\item {\em BB84:} \newline In the BB84 protocol \cite{bb84} Alice
 sends randomly one of the four quantum states \bea \ket{0} & & \ \ ,
 \nonumber \\ \ket{1} & & \ \ , \nonumber \\ \ket{\bar 0} & = & \shalf
 (\ket{0}+\ket{1}) \ \ ,\nonumber \\ \ket{\bar 1} & = & \shalf
 (\ket{0}-\ket{1})\ \ , \label{states} \eea with equal probability.
 Here the states $\ket{0}$ and $\ket{\bar 0}$ represent bit value `0',
 the states $\ket{1}$ and $\ket{\bar 1}$ stand for bit value `1'.  The
 first two states in equation (\ref{states}) correspond to a
 spin-\half -particle being polarized in positive or negative
 $z$-direction, the last two to polarization in positive or negative
 $x$-direction.  This can be graphically visualized as in figure
 \ref{figbb84}.  (All figures in connection with the protocols show
 directions corresponding to polarization vectors of spin-\half
 -particles.)
\newline
\begin{figure}[!ht]
\vspace*{0.2cm}
\centerline{\psfig{width=13cm,file=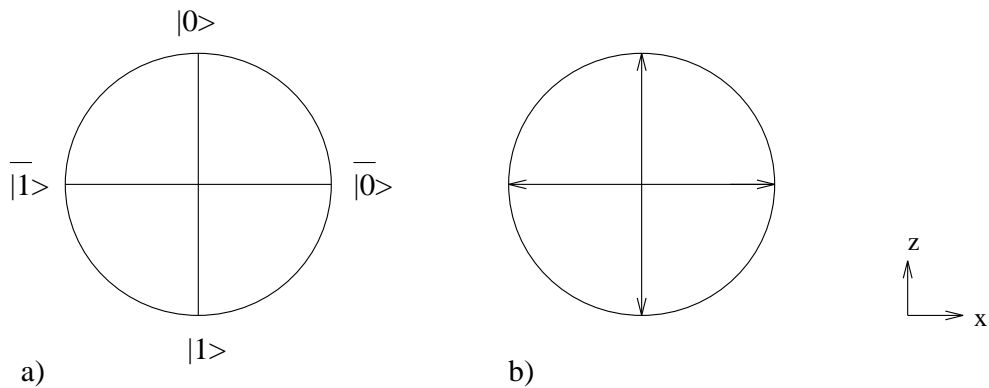}}
\vspace*{5mm}
\caption{ Directions corresponding to polarization of
      a spin-\half -particle for the
      BB84 protocol: a) ensemble of states Alice sends, b) Bob's
    directions of measurement. Note that orthogonal states point in opposite 
    directions,  
    see e.g. $\ket{0}$ and $\ket{1}$,
    which point in $+z$ and $ -z$ direction, respectively.}
\label{figbb84}
\end{figure}
The states in eq. (\ref{states}) can also be represented by linearly
polarized photons: the first two states then correspond to vertically and
horizontally polarized photons, the last two to polarization angles
$45^o$ and $135^o$ with respect to the vertical axis.
\par
When Bob receives a state from Alice, he  
 chooses randomly
 either the $x$- or the $z$-basis for making a measurement. His
result will always be either $\ket{0}$ or $\ket{1}$. But only in the cases 
where he picked the ``right'' basis, i.e. the one which 
Alice used, is his result
correlated with the bit
Alice sent. If, e.g., Alice sent $\ket{\bar 0}$, but Bob measures along the
$z$-direction, he will  find either
$\ket{0}$ or $\ket{1}$ with equal probability. 
After Alice sent and Bob measured the necessary number of states,
 Alice phones Bob (or
uses some other ``classical'' channel) and tells him when she used which basis.
They throw away the cases in which they used different bases,
 and thus have established a
secret key. This key is called the {\em sifted key}.
\par

\item {\em B92:}\newline In this protocol by Bennett \cite{b92} Alice
chooses between two non-orthogonal states to be sent to Bob. It was
shown that in principle {\em any} two non-orthogonal states of a
quantum system can be used for quantum key distribution. Let
$\ket{u_0}$ and $\ket{u_1}$ be the two non-orthogonal states which
represent the bit values 0 and 1, see figure \ref{figb92}.  \par Bob
makes a measurement with a set of so-called POVM's (positive operator
valued measurements), which gives as result either ``$\ket{u_0}$'' or
``$\ket{u_1}$'' or ``I don't know'' (see, e.g., \cite{peres}). For
example, if Alice sends $\ket{u_0}$, Bob will either find $\ket{u_0}$
or an inconclusive result, but never $\ket{u_1}$. They can then use
the public channel to discard inconclusive results, thus arriving at a
correlated string of bits.  \par In practice the two non-orthogonal
states can be realized by two low-intensity coherent states (note that
two different coherent states are never exactly orthogonal, and for
low intensities they become significantly non-orthogonal). An
additional strong reference pulse is used in order to enhance security
of the protocol (see section \ref{problemsand}).
\newline
\begin{figure}[!ht]
\vspace*{0.2cm}
\centerline{\psfig{width=13cm,file=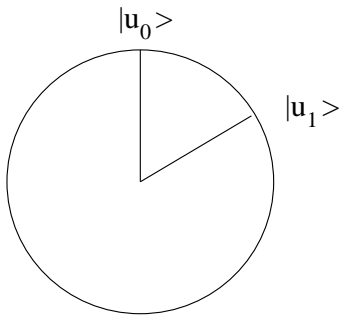}}
\vspace*{5mm}
\caption{ B92 protocol: two non-orthogonal states.}
\label{figb92}
\end{figure}
\item {\em 4+2 protocol:} \newline The protocol described in
\cite{imoto} combines ideas from BB84 and B92: as in BB84 Alice
chooses between two different bases (so the number of possible states
to send is 4), and as in B92 the two states within a basis,
representing bit `0' and `1', are non-orthogonal.  As in B92, a strong
reference pulse is used.  \par Thus, this protocol corresponds to
realizing BB84 with coherent states and a strong reference pulse.
\item {\em Six state protocol:} \newline In the six state protocol
\cite{bruss,gisin} Alice enlarges her ensemble of quantum states she
sends across to Bob, using in addition to the four states in BB84 the
states \bea \ket{\bar{\bar 0}} & = & \shalf (\ket{0}+i\ket{1}) \ \
\text{and} \nonumber \\ \ket{\bar{\bar 1}} & = & \shalf
(\ket{0}-i\ket{1}) \ , \eea which describe a spin-\half -particle
polarized in positive or negative $y$-direction.  (In the case of
photons, these states represent circular polarization.)  The six
states are shown in figure \ref{figsix}. \par Thus, Alice sends a
state randomly polarized in positive or negative $x$-, $y$-, or
$z$-direction to Bob, who measures randomly in the $x-,y-$ or
$z$-basis. As in BB84 they communicate over a public channel and keep
only those cases in which their basis was the same.  \newline
\begin{figure}[!ht]
\vspace*{0.2cm}
\centerline{\psfig{width=13cm,file=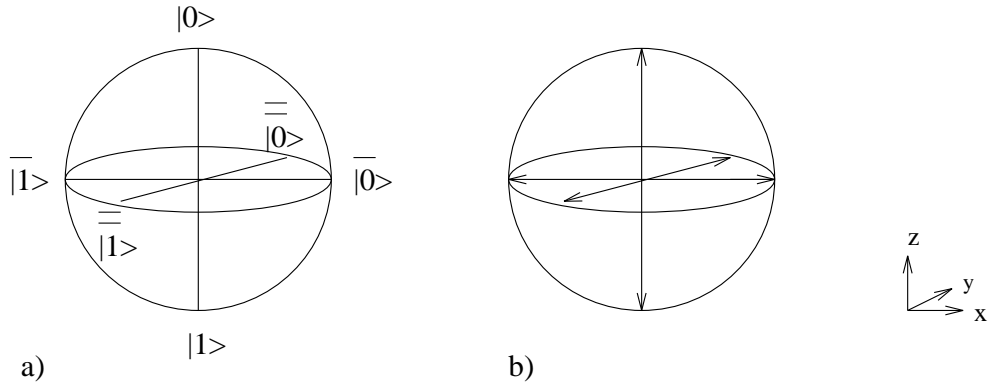}}
\vspace*{5mm} 
\caption{ Six state protocol: a)ensemble of states Alice sends, b)Bob's
    directions of measurement. }
\label{figsix}
\end{figure}

\end{itemize}
\item [2)] {\em Ekert scheme:} \newline
In the key distribution scheme designed by Ekert \cite{ekert}
Alice and Bob are sharing a number of maximally entangled states 
consisting of two two-state systems,  
such that each of them has hold of one of the two correlated systems. 
Let us indicate this by labeling the singlet with indices $A$ and $B$:
\be
\ket{\psi^-} = \shalf (\ket{0}_A\ket{1}_B - \ket{1}_A\ket{0}_B) \ \ .
\ee
They store
their entangled states until they decide to establish the key, then 
Alice chooses
randomly one of the three measurement directions indicated in figure
\ref{figartur}
whereas Bob chooses a set of directions rotated by $45^o$.
\newline
\begin{figure}[!ht]
\vspace*{0.2cm}
\centerline{\psfig{width=13cm,file=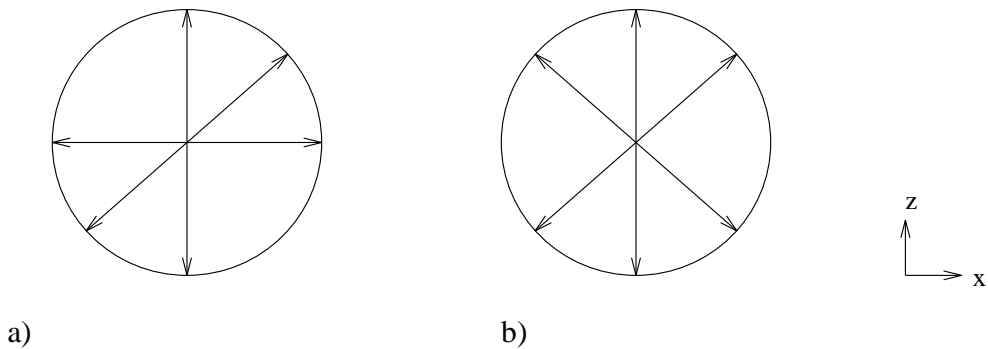}}
\vspace*{5mm}
\caption{ Ekert protocol: a)Alice's directions of
   measurement, b)Bob's
    directions of measurement. }
\label{figartur}
\end{figure}
They use again just those cases in which their measurement directions
were the same. Only then their results are correlated. The runs where
they used different directions can be used to test the Bell inequality
and thus find out whether anybody has interfered with their systems.

\item [3)] {\em Goldenberg/Vaidman class:} \newline
The idea of this class of protocols is to use a superposition of
states, which arrive at different times at Bob's site.
\begin{itemize}
\item {\em Goldenberg/Vaidman:}\newline The scheme described in
\cite{vaidman} uses two orthogonal states, $\ket{\Psi_0}$ and
$\ket{\Psi_1}$, to represent bits `0' and `1', given by \bea
\ket{\Psi_0} & = & \shalf (\ket{a}+\ket{b}) \ \ ,\nonumber \\
\ket{\Psi_1} & = & \shalf (\ket{a}-\ket{b})\ \ ,
\label{ortho}
\eea where $\ket{a}$ and $\ket{b}$ are localized normalized
wavepackets which are sent from Alice to Bob along two channels of
different `length': wavepacket $\ket{b}$ is delayed for some fixed
time until $\ket{a}$ has already reached Bob. This can for example be
achieved by using an interferometer with one short and one long arm.
Bob has to wait with the readout of the superposition until both
$\ket{a}$ and $\ket{b}$ have reached him. In order to make it
impossible for a spy to do her job, the times at which the wavepacket
$\ket{a}$ is sent, have to be random. The advantage of using
orthogonal states is that in principle there is no waste of photons.
\item {\em Koashi/Imoto:} \newline The authors of
\cite{imoto2} show how to circumvent the necessity of random timing by
making the interferometer asymmetric, i.e. by using beamsplitters that
do not have equal transmittivity and reflectivity. This means that the
amplitudes in eq. (\ref{ortho}) change to 
\bea 
\ket{\Psi_0} & = & -i
\sqrt{R} \ket{a}+\ \sqrt{T}\ket{b} \ \ ,\nonumber \\ 
\ket{\Psi_1} & =
& \ \ \ \sqrt{T}\ket{a}-i \sqrt{R}\ket{b}\ \ .
\label{ortho2}
\eea 
The different amplitudes for $\ket{a}$ deprive Eve of the
possibility (given she knows the sending times) to use the simple
strategy to send Bob a dummy $\ket{a}$ and later, after learning the
phase, to send him $\pm \ket{b}$.
 
\end{itemize}
\end{itemize}

\section{Security}
\label{secure}

Due to the principles of quantum mechanics described above,
 it is impossible for the spy Eve to gain
{\em perfect} knowledge of the quantum state sent from Alice to
Bob. Nevertheless, she can acquire {\em some } knowledge. Without
interaction of a spy, each two-level quantum system carries 1 bit of
information (commonly called qubit) from Alice to Bob.  When Eve gets
hold of  part of this information, she cannot prevent causing a
disturbance to the state arriving at Bob's side, and thus introducing
a non-zero error rate.  In principle, Bob can find out about this
error rate and thus about the existence of a spy by communicating with
Alice.  
\par The source for Eve's knowledge are measurements
performed on the  signals (quantum states). 
The simplest eavesdropping attack for Eve
would be to measure each signal just as Bob would do, and then to
resend a signal to Bob which corresponds to the measurement result.
\par However, quantum mechanics allows more general measurements than
these simple projection measurements. Eve can bring an auxiliary
quantum system (a probe) in contact with the signal so that they
interact, and then  perform a projection measurement on
 the auxiliary system to draw some
information about the signal from it. All measurements, including the
simple projection measurements, can be described in this fashion
\cite{helstrom76a,peres}. Another opportunity arises for Eve: she
might delay the measurement of the auxiliary system until she learns
more about the signal during public discussion. An example for useful
information is the signal set from which a signal has been drawn. More
involved strategies within quantum mechanics  
correlate measurements of several signals, thereby attacking the key
as a whole rather than the individual components. This scenario is
referred to as {\em coherent} eavesdropping.  A simpler
class is that of {\em collective} eavesdropping where to each signal
an individual probe is attached just as in the individual
attack. These probes, however, now can be read out together in a
coherent process.

As mentioned above, in the ideal case we are always able to identify an
eavesdropping activity by the occurrence of errors in the
transmission. In a real world this becomes a tricky issue. We will always
 have some detector noise, misalignments of detectors and so
on. It should be pointed out that we cannot even in principle
distinguish errors due to noise from errors due to eavesdropping
activity. We therefore assume that all errors are due to
eavesdropping. An other issue, not discussed here, is that of
statistics. Eavesdroppers can be lucky: they create errors only on
average, so in any specific realization the error rate might be zero
(with probability exponentially small in the key length, of
course). We are guided by the idea that a small error rate, for
example 1 \%, implies that an eavesdropper was not very active, while
a big error rate is the signature of a serious eavesdropping
attempt. But what is the meaning of ``small'' and ``big''?  
\par 
From
an information theoretic point of view, the natural measure of
``knowledge'' about some signal is the Shannon information. It is
measured in bits and can be defined for any two parties, the sender of
the signal and the observer (receiver). In general terms, the
knowledge of the observer consists of obtained measurement results and
any additional gathered knowledge, like the announced basis of signals
in the BB84 protocol. All this knowledge will be denoted by $M$.

From the receiver's point of view there will be an a-priori $p(x)$ and
an a-posteriori $p(x|M)$ probability distribution for the signal
$x$. The knowledge $M$ will turn up with probability $q(M)$. The
Shannon information can now be defined as the {\em expected} change in
entropy of the two probability distributions. It is therefore given by
\be
I = - \sum_x p(x) \log_2 p(x) + \sum_M q(M) \sum_x p(x|M) \log_2 p(x|M) \ \ .
\label{shannon}
\ee
For a binary channel with equal a-priori probabilities for the two
signals the Shannon information can be expressed in terms of the error
probability $e$ with which the signals are received. It is given (in
bits per signal) by
\be
I[e] = 1 + e \log_2 e + (1-e) \log_2 (1-e) \; .
\label{binary}
\ee
 This  is  the Shannon information, 
per element of the sifted
key, between Alice and Bob, $I_{AB}$, with the observed error rate $e$
of the channel. On the other hand we will use the information $I_E$,
 generally given by equation (\ref{shannon}), 
which Eve obtains on the key where $M$ then represents her measurement
results and all the information exchange between Alice and Bob over
the public channel.

Another proposed measure of Eve's knowledge is the probability that
the eavesdropper guesses the correct key given her knowledge about it.

A fundamental difference between classical cryptography and the use of
a one-time pad together with quantum key distribution is that the
former one is vulnerable to technological improvements (faster
computers and algorithms) and  therefore has to be designed to keep the
secret secure against  improvements  which
occur during the whole period of time in which the secrecy is
required. Quantum key distribution, on the other hand, needs to be
designed to be secure only against technology available at the time
(and location) of the quantum part of key distribution. Therefore it
makes sense to give the estimates of the Shannon information for
various scenarios.  They differ by the technology available to
Eve. Examples for potential improvement of Eve's knowledge
are the ability to perform delayed measurements (needs
physical storage of auxiliary quantum systems), the availability of
quantum channels superior to those used by Alice and Bob ( for example
in form of an optical fibre which is less lossy and noisy), and the
ability to perform coherent eavesdropping attacks (needs ability to
manipulate and store coherently several quantum systems).

Let us now quote some results on maximal information leakage to the
eavesdropper. They are valid under the assumption of ideal BB84 signal
states, for example single photons.
\par
It has been shown that the simple intercept-resend strategy
 leads for the BB84 protocol to an average error rate of 25
\% while it yields at best $0.5$ bit of information per
signal \cite{ekert94a,huttner94a}. The optimal probability of a
correct guess would be $75 \%$ in that case.

\begin{figure}[hbt]
\setlength{\unitlength}{1pt}
\centerline{\begin{picture}(250,150)
\epsfysize=5cm
\epsffile[72 230 540 560]{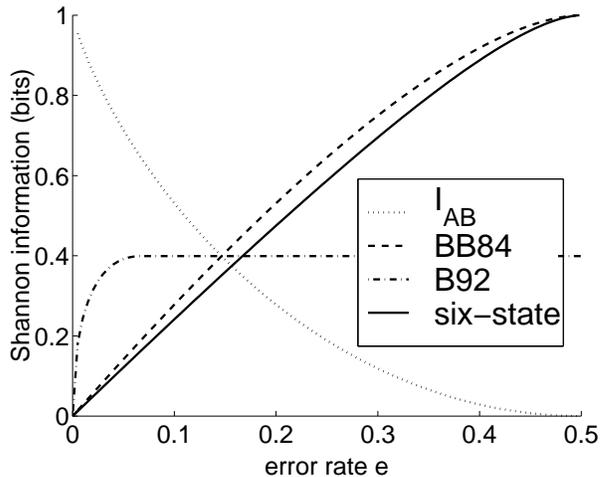}
\vspace{-0.2cm}
\end{picture}}
\vspace{1cm}
\caption[] {\small Maximal mutual information $I_E$ on the sifted key
        shared between Alice and Eve as function of Bob's error rate
        $e$ for the protocols BB84 \cite{fuchs1} and the six-state
        protocol \cite{bruss} together with the the mutual information
        between Alice and Bob given by the curve $I_{AB}$. The graph
        for the B92 protocol \cite{fuchs96a} with state overlap of
        $1/\sqrt{2}$ displays $I_E$ for the raw key and not for the
        sifted key.  }
\label{eave}
\end{figure}

Bounds on the obtainable Shannon information 
for eavesdropping on single bits
can be found in the
literature for different protocols. Fuchs et al. give bounds for the
BB84 \cite{fuchs1} and the B92 protocol
\cite{fuchs96a}. A bound for the six state
protocol was obtained in \cite{bruss}.
 These bounds are illustrated in figure \ref{eave}.
 Note the trade-off between Eve's information gain and 
 the disturbance she causes: more information for Eve means higher 
  error rate for Bob. 
  For reasonably low error rates Eve's maximal information is
  smallest in the six-state protocol, as it uses the biggest ensemble of 
  input states.
 
Bounds for the Shannon information in 
more general attacks are studied in  \cite{cirac} for BB84
and \cite{gisin} for the six-state protocol.

\par The important result from these estimates is that even for
small error rates the eavesdropper might be in possession of
information about the key at a level deemed
dangerous for secure communication. For example, at an observed error
rate of $1 \%$ we find that an eavesdropper might have gained up to
$0.024$ bit of Shannon information per bit of key even for the 
six-state protocol. This is far too high to
allow the direct use of the obtained key for encryption.  Instead, one
uses the tool of privacy amplification \cite{bennett95a} (see
following section) to extract a short secure key from the long
insecure key. 

One of the advantages of the Ekert scheme is that by storing the
states at both ends of the transmission line and coherent manipulation
on each side between the accumulated states the performance of the key
distribution could be enhanced. This technique is called {\em quantum privacy
amplification} \cite{deutsch96a} and effectively gives a new, shorter key
with lower error rate.

\section{Elements for realistic implementations}
\label{elements}
In the previous section we have seen that the Shannon information
available to an eavesdropper about the {\em sifted key} (that is the key
directly after the key exchange) is too high to allow secret
communication directly. Fortunately, it is possible to process this
key with help of a purely classical protocol in order to distill a
new, shorter key from the sifted key which  exponentially
approximates a secret key.  We present the procedure here in a form
which is valid only if Eve's activity is restricted to attacks on
individual signals (as opposed to coherent or collective attacks). 
 However, the steps executed in a quantum key distribution
apparatus are the same in the general case, only the reasoning behind
them changes, as indicated below.

 More details about the full protocol to deal with restricted attacks
 in a realistic scenario can be found in \cite{nl99a}. Here we will
 concentrate only on the main points. An important point for practical
 realization is that in a realistic protocol no ideal public channel
 exists which can be overheard but not changed by an
 eavesdropper. This property of a channel can only be approximated by
 using an open channel where messages will be {\em authenticated} by
 means of a small secret key shared before the start of the
 communication.  Only this method ensures that Alice and Bob do not
 fall victim to the {\em separate world attack}, known as well as {\em
 man in the middle attack}. In this attack an
 eavesdropper cuts the quantum and the classical channel dividing the
 world into two parts. One of these parts 
 contains  Alice, and  Eve
 pretends to her to be Bob and vice versa in the other part.  Alice and Bob
 unknowingly never communicate directly with each other. Only
 authentication by means of previous shared secret knowledge can
 counteract to this attack. In this view quantum key distribution will
 grow a large secret key from a small seed secret key. A by-product of
 this changed scenario is that we are free to use shared secret bits
 in intermediate states to enhance or make clearer the performance of
 the protocol.

The first step in that direction is {\em error correction}. Alice and
Bob exchange redundant  information over the
public channel to reconcile their versions of the key. Obviously, the
amount of exchanged redundant information has to be kept as small as
possible, since the information flow to Eve has to be taken account
of. (One possibility is to encode it using part of the initially
shared secret key.)   What is the minimum amount of
exchanged redundant bits? A correctly received binary string of length
$n_{sif}$ carries exactly $n_{sif}$ bits of Shannon information. On
the other hand, if Bob received this key with an error rate $e$ then
he is in possession of $n_{sif} I_{AB}$ bits only. He, therefore, has to
get hold of the difference of $n_{sif} - n_{sif} I_{AB}[e]$ bits of
information. Since the public channel can be made error free\footnote{Any 
information sent through the public channel can be put into code words, using
any error correction scheme, to protect it against errors. This
encoding into codewords does not change the amount of Shannon
information contained, and one codeword can be regarded as one signal.}
the information per signal sent there is the ideal $1$ bit, so for
each bit of information missing, Alice has to send on average one
signal. Therefore the minimum amount $n_{min}$ of bits to be exchanged
is given by  the Shannon bound,
\be
\label{shannonbound}
n_{min} =  -n_{sif} \left(e \log_2 e + (1-e) \log_2 (1-e)\right) \; .
\ee
 The best known practical protocol is that of
Brassard and Salvail \cite{brassard93a}. It uses an interactive
information exchange between the two sides. The requirements for a
good error correction protocol are to be as close as possible to the
minimum number of exchanged bits given by the Shannon bound and a
success rate of correction as high as possible.  In
contrast to a standard problem in error correction the channel used
for transmission of the redundant bits can be assumed to be error
free, which allows for improved, specialized error correction schemes.

Starting from the reconciled key, Alice and Bob now use {\em privacy
amplification} \cite{bennett95a} to establish a secret key. The idea
behind privacy amplification is to hash the reconciled key of length
$n_{rec}$ into a shorter key of length $n_{fin}$ using random
hashing. An example for hashing is to take parity bits of random
subsets of the reconciled key to form the new key. 
In general,
 we shorten the reconciled key by the fraction $\tau_1$
 and then by additional $n_S$ bits
to a final key length of $n_{fin} = (1-\tau_1) n_{rec} - n_S$.
As shown by Bennett
et al.~\cite{bennett95a}
Eve's  Shannon information on the final key is bounded by
\be
I_E^{final} \leq \log_2 (2^{-n_S} +1) \approx \frac{2^{-n_S}}{\ln 2} \ \ .
\ee
A consequence is that $I_{final}$ can be made exponentially small by
means of the number of security bits $n_S$. 

The central quantity in this context 
is the collision probability $P_{coll}$, and the fraction $\tau_1$
is given by 
$
\tau_1
  =1+\frac{1}{n_{rec}} \log P_{coll}$. 
 Here $P_{coll}$ is
a measure of the a posteriori probability distribution $P_{post}$ of
the reconciled key conditioned on all information available to the
eavesdropper. It is defined by the relation $P_{coll} = \sum_x
\left(P_{post}\right)^2$ where the sum is taken over all reconciled keys.  
For security against eavesdropping
strategies attacking individual signals only it is essential to find
an upper bound on the collision probability. Bounds for $P_{coll}$ and
expressions for $\tau_1$ for the BB84 protocol are given in
\cite{nl96a,slutsky98a,nl99a}, for the B92 protocol in
\cite{slutsky98a} and for the six-state protocol in \cite{gisin}. 

\begin{figure}[hbt]
\setlength{\unitlength}{1pt}
\centerline{\begin{picture}(250,150)
\epsfysize=5cm
\epsffile[72 230 540 560]{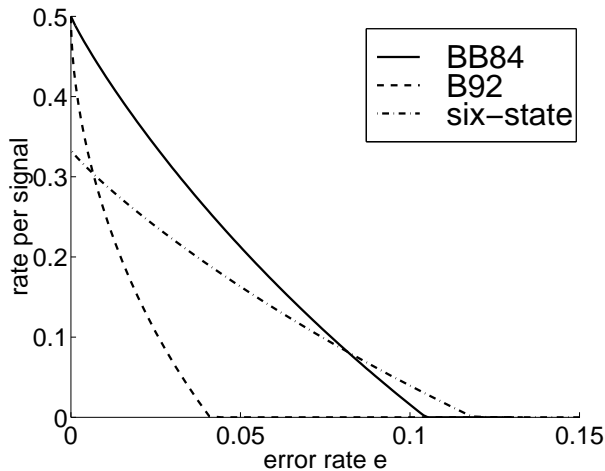}
\vspace{-0.2cm}
\end{picture}}
\vspace{1cm}
\caption[] {\small The rate $\frac{1}{2}R_{del}$ for the B92 protocol with
        overlap $1/\sqrt{2}$ between the two signal states has been
        calculated using the results by Slutsky et
        al.~\cite{slutsky98a}.  The rate $\frac{1}{2}R_{del}$ for the BB84
        protocol is obtained with the estimates from
        \cite{slutsky98a,nl99a} and the estimates leading to the
        rate $\frac{1}{3}R_{del}$ for the six-state protocol are taken from \cite{gisin}.}
\label{rates}
\end{figure}

With these results it is possible to calculate the optimal rate at
which one can extract secure bits from the sifted key. We assume error
correction at the Shannon bound of equation (\ref{shannonbound}) and
encryption of the redundant bits. Then the balance between new secure
bits being created and old secure bits being used up gives an average
creation rate per bit of the sifted key of 
\be R_{corr} = I_{AB}[e] -
\tau_1[e] \ee
 if we use error correction, and 
\be R_{del} = I_{AB}[e]
- \tau_1[e] (1-e) - e \ee
 if we discard errors from the key.  To obtain the creation rate of secure bits as a fraction of the sent quantum signals we have to multiply $R_{corr}$ and $R_{del}$ by a factor $1/2$ for the B92 and the BB84 protocol, and by $1/3$ for the six-state protocol. A direct
comparison for the resulting rates in case of discarded errors  is made
in figure \ref{rates}. The results show that the restriction to
eavesdropping attacks on individual signals allows secure quantum key
distribution with existing experiments. The tolerable error rates, leading to positive rates, are
$4\%$, $10.5 \%$, and $12 \%$ for the three protocols
respectively. The six-state protocol gives the lowest gain for error rates below $\approx 0.65 \%$ while it becomes superior to the BB84 protocol for error rates bigger than approximately $8 \%$.  Though tolerable error rates are
achievable with present day experiments, some work still has to be
done to cope with the signal states which are not the ideal one-photon
states (see section \ref{problemsand}).

For more general strategies than those measuring individual signals 
the
presented way of error correction, estimation of collision
probability, and privacy amplification is no longer valid since in
that case Eve might make use of the knowledge of the particular
hashing function (choice of random subsets for parity bits) to
optimize her measurements. Instead, one has to directly estimate the
Shannon information on the final key. This has been done for a wide
class of  {\em collective attacks} in \cite{biham98a}, while
bounds in the most general case are obtained in
\cite{mayers98a,losub1}. The proof given in \cite{mayers98a} leads to
a maximal tolerated error rate of circa 7 \%. The proof of \cite{losub1}
uses the Ekert scheme in connection with \cite{deutsch96a} to tolerate
higher error rates, as mentioned in the discussion of the Ekert
scheme, but it needs local operations operating with an error rate below the threshold
set for fault tolerant quantum computing.

It is important to note that the key generated by quantum key
distribution is different from the key assumed in the one-time pad or
as seed for the Data Encryption Standard. These keys are assumed to be
absolutely secure and certainly shared between Alice and Bob. The key
established in quantum key distribution does not carry those absolute
attributes. It is not absolute secure. Instead, we can make a
statement about it of the following form: With probability $1-\alpha$
an eavesdropper has less Shannon information than a tolerated value
$I_E^{tol}$ on that key (secrecy) and it is shared between Alice and Bob
with probability $1-\beta$. The two probabilities $\alpha$ and $\beta$
can be made arbitrary small (on cost of the key rate) as long as the
initial error rate is below the cut-off rate mentioned above for the
different scenarios. To our knowledge, this subtle difference between
the key properties assumed in applications and the key properties
resulting from quantum key distribution has not been explored
sufficiently yet. Especially, it would be interesting to explore what
values for $I_E^{tol}$, $\alpha$ and $\beta$ are required for
applications.

\subsection{Problems and practicalities}
\label{problemsand}
All current implementations of quantum key distribution make use of
quantum optical methods. In this context we will discuss realization
issues important for the security aspect without going into technical
details.  The problem of realizing quantum cryptography consists of
three parts: realization of the signal states, transportation of the
signals to Bob and an efficient measurement of the signals.

The simplest choice of signal states, from the theoretical point of
view, are single photons with the polarization as carrier of the
signal. However, at present we do not have a source which would give
us single photons on demand. Instead, one uses weak laser pulses. On
average, each pulse contains typically $0.1$ photons. The photon
number distribution is such that most pulses contain no photon, around
$10 \%$ contain one photon and $1\%$ contain more than one photon. The
pulses containing more than one photon endanger security of
transmission, since an eavesdropper could split off one photon and
extract the full information about the signal later on without causing
any disturbance of the channel. This has to be taken account of when
calculating the amount by which the key is shortened during privacy
amplification. The transmission is totally insecure if the number of
received signals is smaller than the number of multiple-photon signals
sent.

One of the big problems in quantum key distribution is loss of signals
in the fibre. It has been shown that strong loss in the transmission
going together with multi-photon components of the signal states renders key
distribution in all key distribution schemes insecure unless a strong
reference pulse is used \cite{imoto}. This strong reference pulse
is an original part of the B92 protocol \cite{b92}. It fights the problem
that the eavesdropper has means to suppress a signal without causing
errors by sending a vacuum state to Bob. A strong reference pulse,
however, makes sure that no such state exists.

To keep the error rate low, the set-up should be stable under
influence of the environment.  In the case of polarization based
cryptography the main error source is cross talk between the two
polarization modes and a random (classical) rotation of the
polarization along the propagation direction of the fibre.  Here the
proposals of the BB84 or B92 type are easier to implement than the
time separated ideas of Goldenberg/Vaidman and Koashi/Imoto. In the
first group the signal travels from Alice to Bob and is 
influenced by the environment as an entity, 
while in the second group we have two
parts of a signal interacting with two different environments. We
therefore cannot expect the error rates of the second group to be as
good as the $1\%$ error rates of the first group. This is the reason
why no experimental realization of the second group has been tackled
yet.

For the detection schemes we find that it poses a problem to lower the
amplitude of coherent states below a certain point in order to improve
the single-photon approximation. Bob's detectors will give false alarm
(dark counts) with a fixed probability proportional to the time the
detector is gated. Using weaker pulses will increase the number of dark
counts with respect to the real counts, which effectively increases the
error rate because a dark count will give a random measurement result.

One of the advantages of the Ekert scheme is that is allows to use
quantum privacy amplification, thereby giving a new raw key with lower
error rate than the original key. This allows to go below the cut-off
rate for the tolerated error rate even with a noisy channel.  However,
the necessary storing and manipulation devices are not available at
present.

\subsection{Experiments}
Quantum key distribution was implemented for the first time by Bennett
et al. in a demonstration set-up \cite{bennett92a}. The transfer of
the signals took place over $32$ cm of free air with (incoherent)
faint pulses. Experimental demonstrations of the BB84 protocol near to
commercial realizations are reported by the group at British Telecom
by Marand and Townsend \cite{marand95a}. 
Over a  distance up to $30$ km they achieved  an error rate of 
$1.5$--$4$ \% with an average photon number per
pulse of $0.1$--$0.2$ photons. 

Several experiments have been done implementing an approximate B92
protocol. In these experiments the strong reference pulse of the
original scheme is omitted, thereby using the idea of two
non-orthogonal pulses only. It is known that this omission renders the
scheme more insecure.  Best results regarding low error rate are
achieved here by the group in Geneva. They achieve error rates of
about $0.5$--$1.35$ \% over distances of 23 km with an average photon number of
$0.1$--$0.2$ \cite{zbinden98a}. An initial problem of their scheme to give a
low key rate only has now been resolved. Other schemes in free-space
key distribution and over fibre are reported by the Los Alamos group,
going over 40 km in fibre and 1 km in free space \cite{buttler98a}.
Variations of the Ekert scheme have been implemented by Rarity et al.~
\cite{rarity94a}.

\section{Open questions  and summary}
From the point of fundamental physics the most interesting question is
to show security against the most general coherent eavesdropping
attack on single photon signals. This has been achieved
by Mayer \cite{mayers98a} and by Lo and Chau
\cite{losub1}. From the practical point of view these proofs are not
relevant yet since they do not deal with realistic situations. For
this one would need the use of efficient error correction methods, the
ability to cope with large losses and with realistic error rates and,
finally, the extension to realistic signals like dim coherent states
or photons from parametric downconversion.

For practical purposes it makes sense to restrict eavesdropping
strategies to attacks on individual signals. For this scenario
workable schemes for single-photon states have been presented in
\cite{slutsky98a,nl99a}. The extension to realistic signal states has
been achieved recently  \cite{nlsub99a}.

The experimental groups will have to look for set-ups improving the
rate at which the key is generated. It is essential to keep in mind
that it is not the aim to minimize the error rate but to maximize the
key rate!

The questions directed to the audience dealing with the classical part
of quantum key distribution are:
a) what is  a good goal for the security
of the final key? b) How good does it have to be? (In terms of
$I_E^{tol}$, $\alpha$ and $\beta$ as introduced in section
\ref{elements}.) c) What is the optimal reconciliation protocol in these
circumstances?

In summary, quantum key
distribution is a truly interdisciplinary topic in quantum
information. It brings together cryptologists, classical information
scientists, and experimental and theoretical physicists. At present,
there are physical systems which already produce sifted keys at a
reasonable rate with a low error rate. 
Although the implementation is
not ideal, theoretical work should soon show in which scenario it is
possible to extract secure keys from that. To optimize procedures more
work in error correction etc. is needed. After realizing the nature of
security of the final key, we  need more input about the specific
requirements for applications  - as quantum key distribution 
has already passed the first threshold towards implementation.

\section*{Acknowledgments}
The authors took benefit from  the 1998 quantum information workshops at ISI 
(Italy) and Benasque Center for Physics (Spain) and wish to thank their 
organizers and Elsag--Bailey for support.
DB acknowledges support by    the European TMR Research Network 
ERP-4061PL95-1412
and by Deutsche Forschungsgemeinschaft
under grant SFB 407, and  NL by the Academy of Finland.

\bibliography{/home/pcu/hip/lutkenha/tex/STYLES_ETC/norbert} 
\bibliographystyle{/home/pcu/hip/lutkenha/tex/STYLES_ETC/osa}

\end{document}